\documentclass[amsmath,prb,twocolumn,showpacs,floatfix]{revtex4}
\usepackage{graphicx}
\usepackage{amssymb}
\newfont{\Sc}{eusm10}

\begin{document}
\title{Variational speed selection for the interface propagation in superconductors}

\author{Artorix de la Cruz de O\~{n}a}
\email{Artorix.DelaCruz@nbed.nb.ca} \affiliation{A. Center of
theoretical Physics and Applied Mathematics, Dynamical System
Project, Montr\'eal, H3G 1M8, Canada.\\ District Scolaire 9, 3376
rue Principale C.P. 3668, NB E1X 1G5, Canada.}

\date{\today}
%
\vspace{-15mm}
\begin{abstract}
We study the interface propagation in superconductors by means of a
variational method. We compute the lower and upper bounds for which
the planar front speed propagation is valid. To take into account
delay or memory effects in the front propagation, an hyperbolic
differential equation is introduced as an extension of the model.

\end{abstract}
\vspace{5mm}
\maketitle
In recently years the magnetic field penetration or its expulsion
from Superconducting samples have attracted the attention of
different research groups. The understanding of the different
particularities of this phenomenon has been a major challenge.

In Ref.\onlinecite{barto}, the authors have obtained the interface
speed by using heuristic methods such as Marginal stability
hypothesis(MSH) and Reduction order.

In this paper, we aboard the interface propagation speed from the
variational point of view. The variational speed selection or BD
method was proposed by Benguria and
Depassier\cite{Bengu1,Bengu2,Bengu3} in order to study the
reaction-diffusion equations. Using a trail function $g(x)$ in the
procedure, one may find accurate lower and upper bounds for the
speed $c$. The function $g(x)$ must satisfy that $g(x)>0$ and
$g'(x)<0$ in $(0,1)$. Only if the lower and the upper bounds
coincide can the value of $c$ be determined without any uncertainty.

Our start point are the Ginzburg-Landau equations\cite{dorsey},
which comprise a coupled equations for the density of
superconducting electrons and the local magnetic field. In order to
describe the evolution of the system between two homogeneous steady
states, we assume a SC sample embedded in a stationary applied
magnetic field equal to the critical $H=H_{c}$. The magnetic field
is rapidly removed, so the unstable normal-superconducting planar
interface propagates toward the normal phase so as to expel any
trapped magnetic flux, leaving the sample in Meissner state. Also,
we have considered that the interface remains planar during all the
process.

To take into account the delay effect in the interface propagation,
due to, for example, imperfections and nonhomogeneous
superconducting properties in the material, we have included the
delay time $\tau$ and indeed introduce the hyperbolic
differential(HD) equation. This type of equation has been recently
applied in biophysics to model the spread of humans\cite{fort},
bistable systems\cite{mendez1}, forest fires\cite{mendez2} and in
population dynamics\cite{mendez3}.
%

\emph{Traveling wave solutions}. We are interested in finding
traveling wave solutions for our model. To start we use the
one-dimensional time-dependent Ginzburg-Landau equations(TDGL),
which in dimensionless units\cite{dorsey} are
\begin{eqnarray}
\label{eq:ginzburg}
\partial_{t}f=\frac{1}{\kappa^2}\,\,\partial^{2}_{x}f-q^{2}f+f-f^{3}\nonumber,
\\
\bar{\sigma}\partial_{t}q=\partial^{2}_{x}q-f^{2}q,
\end{eqnarray}
where $f$ is the magnitude of the superconducting order parameter,
$q$ is the gauge-invariant vector potential (such that
$h=\partial_{t}q$ is the magnetic field), $\bar{\sigma}$ is the
dimensionless normal state conductivity (the ratio of the order
parameter diffusion constant to the magnetic field diffusion
constant) and $\kappa$ is the Ginzburg-Landau parameter which
determines the type of superconducting material; $\kappa<1/\sqrt{2}$
describes what are known as type-I superconductors, while
$\kappa>1/\sqrt{2}$ describes what are known as type-II
superconductors.

In our analysis we will search for steady traveling waves solutions
of the TDGL equations of the form $f(x,t)=s(x-c\,t)$ and
$q(x,t)=n(x-c\,t)$, where $z=x-c\,t$ with $c>0$. Then the equations
become
\begin{eqnarray}
\label{eq:steady_equation}
\frac{1}{\kappa^2}\,\,s_{zz}+c\,s_{z}-n^{2}s+s-s^{3}=0\nonumber,
\\
n_{zz}+\bar{\sigma}c\,n_{z}-s^{2}n=0,
\end{eqnarray}
\section{Variational analysis}
\emph{Vector potential $q=0$}. In this section, we assume $q=0$ for
the TDGL equations,
\begin{eqnarray}
\label{eq:newginzburg}
\partial_{t}f=\frac{1}{\kappa^2}\,\,\partial^{2}_{x}f+f-f^{3}.
\end{eqnarray}
Then, there exists a front $f=s(x-ct)$ joining $f=1$, the state
corresponding to the whole superconducting phase to $f=0$ the state
corresponding to the normal phase. Both states may be connected by a
traveling front with speed $c$. The front satisfies the boundary
conditions $\lim_{s\rightarrow-\infty} f=1,\lim_{s\rightarrow\infty}
f=0$. Then Eq.(\ref{eq:newginzburg}) can be written as,
\begin{eqnarray}
\label{eq:transf}
s_{zz}+c\,\kappa^{2}\,s_{z}+\mathfrak{F}_{k}(s)=0,
\end{eqnarray}
where $\mathfrak{F}_{k}$ is given by $\mathfrak{F}_{k} =
\kappa^{2}\,s(1-s^{2})$.

We define $p(s)=-ds/dz$, where the minus sign is included so that
$p$ is positive. One finds that the front is solution of
\begin{eqnarray}
\label{eq:monot_equat}
p\,(s)\,\frac{dp\,(s)}{ds}-c\,\kappa^{2}\,p\,(s)+\mathfrak{F}_{k}(s)=0,
\end{eqnarray}
with $p\,(0)=0$, $p\,(1)=0$, $p>0$ in $(0,1)$.

Let $g$ be any positive function in $(0,1)$ such that $h=-dg/ds>0$.
Multiplying Eq.(\ref{eq:monot_equat}) by $g(s)$ and integrating by
parts between $s=0$ and $s=1$ and taking into account $hp+(g
\,\mathfrak{F}_{k}/p)\geq 2\,\sqrt{g\,h \,\mathfrak{F}_{k}}$, we
obtain that,
\begin{eqnarray}
\label{eq:varia_veloc}
c\geq\,\frac{2}{\kappa}\,\int^{1}_{0}
(g\,h\,\mathfrak{F})^{\frac{1}{2}}\,\, ds / \int^{1}_{0} g\,\, ds,
\end{eqnarray}
\begin{figure}[!tbp]
\centerline{
\includegraphics[width=3.2in]{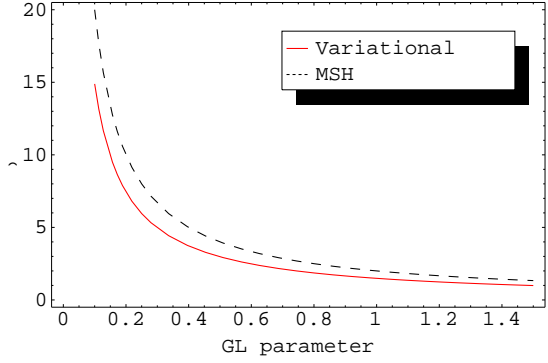}
}\caption{Predictions for the speed. The continuous curve is the
lower bound predicted by the BD method given in
Eq.(\ref{eq:veloci_valuee}). The dashed curve is for MSH.}
\label{figure1}
\end{figure}
As a trial function we have chosen $g(s)=(1-s)^{2}$. Then one finds
that,
\begin{eqnarray}
\label{eq:veloci_integral}
c\geq\,\frac{2}{\kappa}\,\frac{\left[\int^{1}_{0}
s(1-s)^{2}(1-s^{2})(2-2s)\right]^{1/2} ds} {\int^{1}_{0}
(1-s)^{2}\,ds}.
\end{eqnarray}
after integration the speed is given by,
\begin{eqnarray}
\label{eq:veloci_valuee}
c\,\geq\,\frac{3}{64\,k}\,
\left[124+37\sqrt{2}\,\log(3-2\,\sqrt{2})\right].
\end{eqnarray}
Notice that $c\,\leq\,2/\kappa,$where $2/\kappa$ is the result
obtained by using the MSH method. In Fig.1, the graphic shows that
for values $\kappa>1.4$ the MSH speed tends to the BD value, but for
$\kappa<1.4$ the variational speed selection provides a better lower
bound.

\emph{Vector potential $q=1-f$}. For a set of parameters\cite{barto}
$\kappa=1/\sqrt{2}$ and $\bar{\sigma}=1/2$, we have that
$s(z)+n(z)=1$, then Eq.(\ref{eq:steady_equation}) takes the form,
\begin{eqnarray}
\label{eq:par_eq}
s_{zz}+\frac{c}{2}\,s_{z}+\mathfrak{F}(s)=0,
\end{eqnarray}

With this in mind, we look for solutions of the form $s(z)=1-n(z)$.
Proceeding as in Eq.(\ref{eq:varia_veloc}) we have that,
\begin{eqnarray}
\label{eq:velocinte}
c\geq\,2\,\sqrt{2}\,\int^{1}_{0}
(g\,h\,\mathfrak{F})^{\frac{1}{2}}\,\, ds / \int^{1}_{0} g\,\, ds,
\end{eqnarray}
then,
\begin{eqnarray}
\label{eq:veloci_integral}
c\geq\,2\,\sqrt{2}\,\,\frac{\int^{1}_{0}\,\left[n^{2}(1-n)^{2}(1-n)(2-2n)\right]^{1/2}
ds}{\int^{1}_{0} (1-n)^{2}\,ds},
\end{eqnarray}

Finally, for the Eq.(\ref{eq:veloci_integral}) we arrive to
$c\,\geq\,1,$ which is a better lower bound than the $\sqrt{2}$
predicted by the MSH method.
\section{Front flux expulsion with delay}
An import feature phenomena is the existence of a delay time. In
systems with interface propagation, this can be taken into account
by resorting to the hyperbolic differential equation seen in Section
I, which generalizes the parabolic equation. The aim of this section
is to study the interface speed problem in superconducting samples
by means of the HD equations.

Our starting point is the HD equation,
\begin{eqnarray}
\label{eq:generalHRD}
\tau\,\frac{\partial^2\,u}{\partial\,t^{2}}+\frac{\partial\,u}{\partial\,t}=\frac{\partial^2\,u}{\partial\,x^{2}}
+f(u)+\tau\,\frac{\partial\,f(u)}{\partial\,t}.
\end{eqnarray}
In the absence of a delay time $(\tau =0)$, this reduces to the
classical equation $u_{t}=u_{xx}+f(u)$.

\emph{Vector potential $q=0$}. Taking into account the
Eqs.(\ref{eq:ginzburg}) and (\ref{eq:generalHRD}) we can write the
following expression,
\begin{eqnarray}
\label{eq:newgeneralHRD}
\kappa^{2}\,\tau\,\frac{\partial^2\,f}{\partial\,t^{2}}+\kappa^{2}\,\frac{\partial\,f}{\partial\,t}=\frac{\partial^2\,f}{\partial\,x^{2}}
+\kappa^{2}\,\mathfrak{F}+\kappa^{2}\,\tau\,\frac{\partial\,\mathfrak{F}}{\partial\,t},
\end{eqnarray}
where $\mathfrak{F} = s(1-s^{2})$.

It has been proved\cite{mendez1,mendez2,mendez3} that
Eq.(\ref{eq:generalHRD}) has traveling wave fronts with profile
$s(x-ct)$ and moving with speed $c>0$. Then we can write
Eq.(\ref{eq:newgeneralHRD}) as follows,
\begin{eqnarray}
\label{eq:transHRDd}
(1-a\,c^{2})\,s_{zz}+c\,[\kappa^{2}-a\,\mathfrak{F}'(s)]\,s_{z}+\mathfrak{F}_{k}(s)=0,
\end{eqnarray}
where $z=x-ct$, $a=\kappa^{2}\,\tau$, $\mathfrak{F}_{k} =
\kappa^{2}\,\mathfrak{F}$, and with boundary conditions
$lim_{z\rightarrow\infty}s=0$,\,$lim_{z\rightarrow-\infty}s=1$, and
$s_{z}<0$ in $(0,1)$; $s_{z}$ vanishes for $z\rightarrow\pm\infty$.

For the variational analysis we define $p\,(s)=-s_{z}$ with
$p\,(0)=p\,(1)=0$ and $p>0$ in $(0,1)$. Then the
Eq.(\ref{eq:transHRDd}) may be written as
\begin{eqnarray}
\label{eq:newtransHRDd}
(1-a\,c^{2})\,p\,\frac{dp}{ds}-c\,[\kappa^{2}-a\,\mathfrak{F}'(s)]\,p+\mathfrak{F}_{k}(s)=0.
\end{eqnarray}
Multiplying Eq.(\ref{eq:newtransHRDd}) by $g/p$ where $g$ is an
arbitrary positive function and integrating by parts, we have that
\begin{eqnarray}
\label{eq:partintegration}
c\,\kappa^{2}\int_{0}^{1}g[1-\frac{a}{\kappa^{2}}\,\mathfrak{F}']ds=
\int_{0}^{1}[(1-a\,c^{2})hp+\frac{g \mathfrak{F}_{k}}{p}]ds
\end{eqnarray}
where we have used the relation
\begin{eqnarray}
\label{eq:relation}
(1-a\,c^{2})hp+\frac{g \,\mathfrak{F}_{k}}{p}\geq
2\sqrt{1-ac^{2}}\,\sqrt{g\,h \,\mathfrak{F}_{k}}\,\,,
\end{eqnarray}
and $h=-g'>0$.
\begin{eqnarray}
\label{eq:newveloc}
\frac{c}{\sqrt{1-a\,c^{2}}}\geq\,2\,\kappa\,\frac{\int^{1}_{0}
(g\,h\,\mathfrak{F})^{1/2}\,ds}{\int^{1}_{0}
g(\,\kappa^{2}-a\,\mathfrak{F}')\,ds}.
\end{eqnarray}
The maximum is attained for a $g$. Thus, the expression for the
velocity is given by
\begin{eqnarray}
\label{eq:newuppera}
c\,\geq\,2\kappa\,\frac{I_{1}}{\left[I^{2}_{2}\,+\,4\,\kappa^{2}\,a\,I_{1}^{2}\right]^\frac{1}{2}},
\end{eqnarray}
\begin{eqnarray}
\label{eq:ione}
I_{1}\equiv\int^{1}_{0}
\sqrt{gh\mathfrak{F}}ds,\,\,\,I_{2}\equiv\int^{1}_{0}
g(\kappa^{2}-a\mathfrak{F}') ds,
\end{eqnarray}
Notice that if the delay time is neglected $a=0$, this reduces to
Eq.(\ref{eq:varia_veloc}).

\emph{The lower bound}. To compute the lower bound we start with the
trial function given by $g(s)=(1-s)^{2}$ and the expression for
$\mathfrak{F}_{k}$, which both are substituted in
Eq.(\ref{eq:newuppera}). Then,
\begin{eqnarray}
\label{eq:ionedevelop}
I_{1}=\int_{0}^{1}\left[2n(n^{2}-1)(n-1)^{3}\right]^{\frac{1}{2}}\,dn,\nonumber\\
I_{2}=\int_{0}^{1}(1-n)^{2}\left[\kappa^{2}-a\,(1-3\,n^{2})\right]\,dn.
\end{eqnarray}
from Eq.(\ref{eq:newuppera}) we have that,
\begin{eqnarray}
\label{eq:newupperaresult}
c\,\geq\,2\kappa\,\frac{\mathcal{J}}{\left[1\,+\,4\,\kappa^{2}\,a\,\mathcal{J}^{2}\,\right]^\frac{1}{2}},
\end{eqnarray}
where
\begin{eqnarray}
\label{eq:result}
\mathcal{J}=\frac{15\left[124+37\sqrt{2}\,\log\left(3-2\sqrt{2}\right)\right]}{64(10k^{2}-7\,a)}.
\end{eqnarray}

\emph{The upper bound}. The upper bound can be computed by using the
Jensen's inequality\cite{Bengu2},
\begin{eqnarray}
\label{eq:inversejensen}
\frac{\int^{1}_{0} \mu(s)\,\sqrt{\alpha(s)}\,ds}{\int^{1}_{0}
\mu(s)\,ds}\leq\,\sqrt{\frac{\int^{1}_{0}
\mu(s)\,\alpha(s)\,ds}{\int^{1}_{0} \mu(s)\,ds}},
\end{eqnarray}
where $\mu(s)>0$ and $\alpha(s)\geq0$. We define $\mu(s)=
g(\kappa^{2}-a\,\mathfrak{F}')$ and
$\alpha(s)=\mathfrak{F}\,h/g(\kappa^{2}-a\,\mathfrak{F}')^2$. Then
we can write
\begin{eqnarray}
\label{eq:upper_comp}
\frac{\int^{1}_{0} (g\,h\,\mathfrak{F})^{1/2}\,ds}{\int^{1}_{0}
g(\,\kappa^{2}-a\,\mathfrak{F}')\,ds}\,\leq\,\left[
\frac{\int^{1}_{0}
\left[h\,\mathfrak{F}/(\kappa^{2}-a\,\mathfrak{F}')\right]\,ds}{\int^{1}_{0}
g\,(\kappa^{2}-a\,\mathfrak{F}')\, ds}\right]^{1/2},
\end{eqnarray}
where
\begin{eqnarray}
\label{eq:divupperfirst}
\int^{1}_{0}
\frac{h\mathfrak{F}}{(\kappa^{2}-a\,\mathfrak{F}')}ds=\int^{1}_{0}\frac{
n[1-n^{2}(2-2n)]}{[\kappa^{2}-a(1-3n^{2})]}dn,\nonumber\\
\int^{1}_{0} g(\kappa^{2}-a\mathfrak{F}') ds=\int^{1}_{0}
(1-n)^{2}[\kappa^{2}-a(1-3n^{2})]dn,
\end{eqnarray}
then we have that,
\begin{figure}[!tbp]
\centerline{
\includegraphics[width=3.2in]{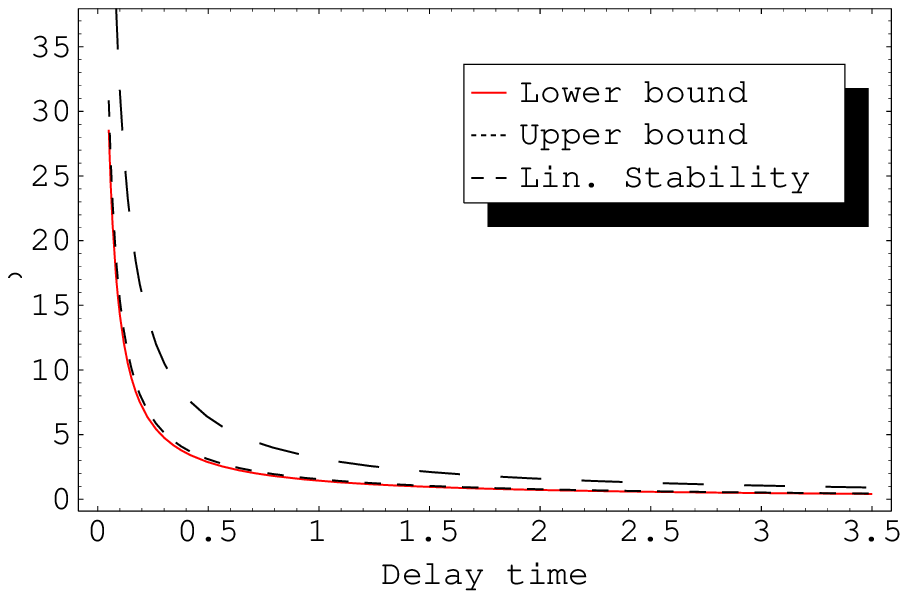}
}\caption{Time-delayed interface propagation speed for $q=0$. The
plot of the lower and upper bounds from variational method as well
as the speed proposed by linear stability are shown.}
\label{figure2}
\end{figure}
\begin{eqnarray}
\label{eq:newres}
c\,\leq\,2\kappa\,\frac{\mathcal{I}}{\left[1\,+\,4\,\kappa^{2}\,a\,\mathcal{I}^{2}\,\right]^\frac{1}{2}},
\end{eqnarray}
where,
\begin{eqnarray}
\label{eq:addition}
\mathcal{I}=(1/3)\,(l_{1}+l_{2})^{\frac{1}{2}},
\end{eqnarray}
and,
\begin{eqnarray}
\label{eq:additionfirst}
l_{1}=\frac{1}{a^{2}}\left[(2a+\kappa^{2})\,\log\left(\frac{\kappa^{2}-a}{2a+\kappa^{2}}\right)-(5a+2\kappa^{2})\right],\nonumber\\
l_{2}=\left[\frac{2(\kappa^{2}-a)^\frac{1}{2}(2a+\kappa^{2})\arctan\left(\sqrt{\frac{3a}{\kappa^{2}-a}}\right)}{\sqrt{3}\,a^{5/2}}\right].
\end{eqnarray}
The Eq.(\ref{eq:newres}) gives a better upper bound than the one
predicted by linear stability\cite{mendez1} i.e,
$c<c_{max}=1/\sqrt{a}$.

In Fig.2 we have plotted the results of the BD method given by
Eqs.(\ref{eq:newupperaresult}) and (\ref{eq:newres}) as well as the
bound proposed by linear stability(LS) methodology. The interface
speed propagation can be predicted in a precisely way by using our
trial function. On the other hand the difference with linear
stability result is notable.

\emph{Vector potential $q=1-f$}. Taking into account the
Eqs.(\ref{eq:ginzburg}) and (\ref{eq:generalHRD}) we can write the
following expression,
\begin{eqnarray}
\label{eq:secondgeneralHRD}
\frac{\tau}{2}\,\frac{\partial^2\,f}{\partial\,t^{2}}+\frac{1}{2}\,\frac{\partial\,f}{\partial\,t}=\frac{\partial^2\,f}{\partial\,x^{2}}
+\frac{1}{2}\,\mathfrak{F}+\frac{\tau}{2}\,\frac{\partial\,\mathfrak{F}}{\partial\,t},
\end{eqnarray}
where $\mathfrak{F} = s^{2}(1-s)$.

Then we can write Eq.(\ref{eq:secondgeneralHRD}) as follows,
\begin{eqnarray}
\label{eq:transHRD}
(1-a\,c^{2})\,s_{zz}+c\,[\kappa^{2}-a\,\mathfrak{F}'(s)]\,s_{z}+\mathfrak{F}_{k}(s)=0,
\end{eqnarray}
where we have assumed $\mathfrak{F}_{k}=(1/2)\mathfrak{F}$ and
$a=\tau/2$.

The expression for the velocity is given by
\begin{eqnarray}
\label{eq:newuppers}
\frac{c}{\sqrt{1-ac^{2}}}\geq\,2\,\sqrt{2}\,\frac{\int^{1}_{0}
(g\,h\,\mathfrak{F})^{1/2}\,ds}{\int^{1}_{0}
g(1-2\,a\,\mathfrak{F}')\, ds}.
\end{eqnarray}

Proceeding as in Eq.(\ref{eq:newveloc}), we get the following
expression,
\begin{eqnarray}
\label{eq:speedlow}
c\,\geq\,2\,\sqrt{2}\,\,\,\frac{\mathfrak{I}_{1}}{\left(\mathfrak{I}^{2}_{2}\,+\,8\,a\,\mathfrak{I}_{1}^{2}\right)^{1/2}}
\end{eqnarray}
where
\begin{eqnarray}
\label{eq:again}
\mathfrak{I}_{1}\equiv\int^{1}_{0}
\sqrt{gh\mathfrak{F}}ds,\,\,\,\mathfrak{I}_{2}\equiv\int^{1}_{0}
g(1-2\,a\,\mathfrak{F}') ds,
\end{eqnarray}
\emph{The lower bound}. As mentioned before, one may obtain lower
bound for the interface speed by means of our trial function $g(n)$.
Taking into account Eqs.(\ref{eq:again}), the integral functions can
be written as,
\begin{eqnarray}
\label{eq:developed}
\mathfrak{I}_{1}=\int_{0}^{1}\left[n^{2}(2-2n)(1-n)^{3}\right]^{\frac{1}{2}}\,dn,\nonumber\\
\mathfrak{I}_{2}=\int_{0}^{1}(1-n)^{2}\left[1-2\,a\,(2\,n-3\,n^{2})\right]\,dn.
\end{eqnarray}

Then the velocity takes the form,
\begin{eqnarray}
\label{eq:veloc_lower}
c\,\geq\,\sqrt{5}\,\,(15-a)^{-\,1/2},
\end{eqnarray}

\emph{The upper bound}. To compute the upper bound we have used the
expression Eq.(\ref{eq:upper_comp}) but with
$\mathfrak{F}=s^{2}(1-s)$ and $\kappa^{2}=1/2$, then
\begin{eqnarray}
\label{eq:velocnewuppers}
\frac{c}{\sqrt{1-ac^{2}}}\leq\,2\,\sqrt{2}\,\frac{\int^{1}_{0}
\left[h\,\mathfrak{F}/(1-2\,a\,\mathfrak{F}')\right]\,ds}{\int^{1}_{0}
g(1-2\,a\,\mathfrak{F}')\, ds}.
\end{eqnarray}
\begin{eqnarray}
\label{eq:deve}
\int^{1}_{0}
\frac{h\,\mathfrak{F}}{(1-2\,a\,\mathfrak{F}')}\,ds=\int_{0}^{1}\left[n^{2}(2-2n)(1-n)\right]^{\frac{1}{2}}\,dn,\nonumber\\
\int^{1}_{0} g(1-2\,a\,\mathfrak{F}')\,
ds=\int_{0}^{1}(1-n)^{2}\left[1-2\,a\,(2\,n-3\,n^{2})\right]\,dn.
\end{eqnarray}
\begin{figure}[!tbp]
\centerline{
\includegraphics[width=3.2in]{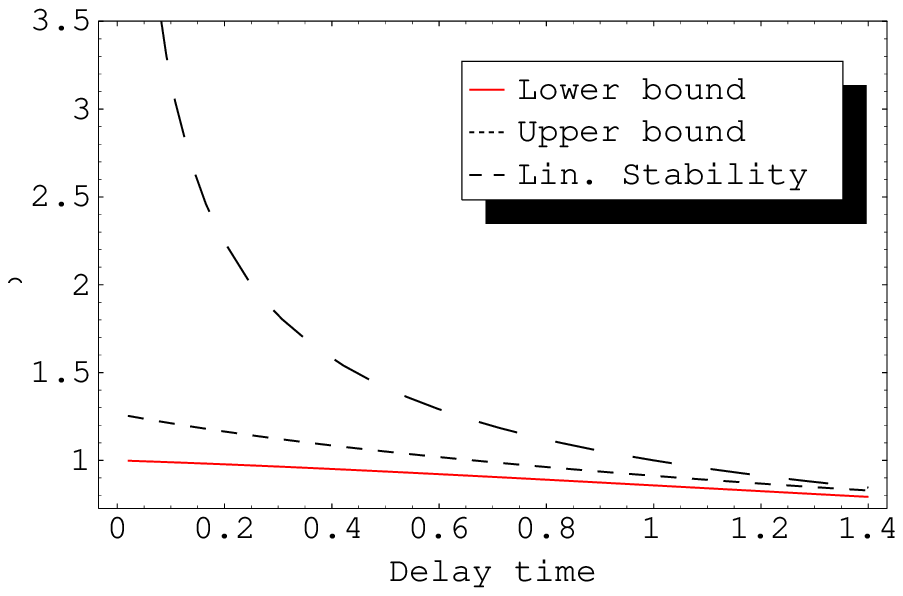}
}\caption{Time-delayed interface propagation for $q=1-f$. The plot
of the lower and upper bounds from variational method as well as the
speed proposed by linear stability are shown.} \label{figure3}
\end{figure}
After integrating and do some algebra, the expression for the
velocity is given by
\begin{eqnarray}
\label{eq:uspeedlow}
c\,\geq\,2\,\sqrt{2}\,\,\,\frac{\mathfrak{B}}{\left[(\frac{1}{3}-\frac{2\,a}{15})\,
+\,8\,a\,\mathfrak{B}^{2}\right]^{1/2}},
\end{eqnarray}
where,
\begin{eqnarray}
\label{eq:B1}
\mathfrak{B}=\frac{\beta_{1}}{18\sqrt{a^{5/2}}}\left(\frac{2\arctan\alpha}{\sqrt{6-4a}}
+\frac{\arctan4\alpha}{\sqrt{\frac{3}{2}-a}}+\beta_{2}\right)^{1/2},
\end{eqnarray}
\begin{eqnarray}
\label{eq:B11}
\alpha\equiv\,\sqrt{a\left(\frac{3}{2}-a\right)}/(3-a),
\\\nonumber \beta_{1}\equiv\,8a^{2}+6a+9,\\\nonumber
\beta_{2}\equiv\,2\sqrt{a}\left[-3(3+4a)+(3+2a)\log(1+2a)\right].
\end{eqnarray}
In Fig.3 we have plotted the results of the BD method given by
Eqs.(\ref{eq:veloc_lower}) and (\ref{eq:uspeedlow}) as well as the
bound proposed by LS method. The interface speed propagation can be
predicted in a precisely way by using this trial function. On the
other hand the difference with LS result is notable.

\emph{Conclusion}. Throughout this work, we have performed
analytical analyses on the superconducting-normal interface
propagation speed problem in parabolic and hyperbolic equations. We
have made use of the variational analysis to obtain the lower and
upper bounds for the speed in each case.

\end{document}